\documentclass[intlimits,twoside,a4paper]{article}

\usepackage[utf8]{inputenc}

\usepackage{physics}
\usepackage{subcaption}
\usepackage{booktabs}

\usepackage[normalem]{ulem}
\usepackage{graphicx}
\usepackage{eso-pic}
\usepackage{xcolor}

\usepackage[eqsecnum]{cmpj3}

\issue{2026}{29}{3}{33603}
\doinumber{10.5488/CMP.29.33603}

\title[Phase diagram of a double-occupancy cell model]%
{Phase diagram of a double-occupancy cell model of a fluid with Curie-Weiss interaction}
\author[R. V. Romanik, O. A. Dobush, M. P. Kozlovskii, I. V. Pylyuk, M. A. Shpot]{R.~V.~Romanik\orcid{0009-0008-2405-9038}\thanks{Corresponding author: \email{romanik@icmp.lviv.ua}.},
	O.~A.~Dobush\orcid{0000-0001-9237-092X},
	M.~P.~Kozlovskii\orcid{0000-0002-8170-4215},
	I.~V.~Pylyuk\orcid{0000-0003-1982-933X},
	M.~A.~Shpot\orcid{0000-0001-9098-9745}}
\address{Yukhnovskii Institute for Condensed Matter Physics of the National Academy of Sciences of Ukraine, 1~Svientsitskii~Str., 79011 Lviv, Ukraine}

\date{Received 3 June 2026; revised 14 August 2026; accepted 31 August 2026; published 28 September 2026}

\begin{document}

\maketitle

\begin{abstract}
A double-occupancy cell model of a fluid with Curie-Weiss interaction is studied. First, we show that the model is isomorphic to the Blume-Capel model on a complete graph through a simple transformation from spin to occupancy variables. We then investigate its phase behavior within the grand-canonical ensemble using a combination of analytical and numerical methods. Despite its simplicity, the model exhibits a remarkably rich thermodynamic behavior depending on the ratio between the local repulsive and global attractive interactions. We identify the regimes characterized by a single critical point, two distinct critical points, tricritical behavior, and triple-point formation. For sufficiently strong repulsion, the system possesses three fluid phases of different densities, leading to both gas-liquid and liquid-liquid coexistence. The locations of the critical, tricritical, and triple points are determined, and the corresponding phase diagrams are constructed. These results demonstrate that the competition between double-occupancy repulsion and long-range attraction is sufficient to generate a complex phase behavior in a minimal multiple-occupancy lattice-gas-like model.
%
%
\printkeywords
\end{abstract}

	\section{Introduction}\label{sec:intro}

Soft matter physics extensively relies on the soft-core, or bounded, interaction potentials~\cite{Likos01}. They are considered as effective potentials for interactions between complex macromolecules in solutions. Examples are the solutions of polymer chains~\cite{LBHM00,BLHM01}, dendrimers~\cite{LSLetal01,MKL08}, star polymers~\cite{LLWetal98}, and other so-called ultrasoft colloids~\cite{Likos06}. The effect of interpenetrability of molecules leads to interesting physical phenomena, not observed in simple fluids with hard-core short-range interactions. Examples of such phenomena are the reentrant melting~\cite{LWL98} with the formation of associated high-density fluids~\cite{LLWL00}, as well as the emergence of clustered crystalline phases~\cite{MGKNL06,MCF07,MKL08} and cluster fluids~\cite{LMLKB10}. Phase diagrams of such soft-particle systems can be quite complex, featuring multiple first-order phase transition lines, each terminating at its own critical point~\cite{NL11, WS13, WS14, Prestipino14, PGT15, dMDM23}. In some theoretical models, the number of critical points becomes infinite. The possibility of particle interpenetration in systems with bounded interactions naturally suggests a discrete lattice mapping that allows for multiple occupancy of lattice cells in their theoretical description.
Resulting models, often called ``coarse-grained'' in soft-matter physics, provide a simplified framework for studying thermodynamic effects associated with ultrasoft effective interactions, including clustering phenomena and complex phase behavior.

Lattice gas models have played an important role in understanding different phenomena in condensed matter physics~\cite{Baxter82,FriedliVelenik17}. For efficient description of soft matter systems with bounded interactions, the corresponding lattice models have to allow multiple cell occupancy to represent the interpenetrability of particles. Surprisingly, lattice gases with multiple occupancy have only recently gained more attention due to the development in soft matter physics and applications of bounded effective potentials to describe the interactions between complex macromolecules that can interpenetrate with a finite energy penalty. In particular, models with unrestricted occupancies were studied in \cite{FHL04,FL18,FL20}, and in our previous works~\cite{KKD20,KD22,RDKPS26a}. A particularly simple realization of a multiple-occupancy lattice model is the double-occupancy case, where each cell can contain at most two particles. Such a model was recently studied in~\cite{LYZ21} on a two-dimensional lattice with nearest-neighbor interactions, where it was shown to be isomorphic to the well-known Blume-Capel model~\cite{Blume66,Capel66}. This is intuitively clear because the occupancy variable $n_l$ takes on three values, $n_l\in\{0,1,2\}$, which naturally associates with spin-$1$ values $S_l\in\{-1,0,1\}$.
Moreover, this is in complete analogy with the isomorphism between the well-known single-occupancy lattice gas and the spin-$1/2$ Ising model~\cite{LY52}.

Despite its apparent simplicity, the double-occupancy model represents the simplest nontrivial member of the family of multiple-occupancy lattice gases demonstrating a much richer phase behavior than the conventional single-occupancy lattice gas. Whereas the nearest-neighbor version of the model has already been studied~\cite{LYZ21}, its Curie--Weiss counterpart has not been investigated in detail. In particular, the structure of its phase diagram, the conditions under which multiple critical points emerge, and the possible existence of tricritical and triple points remain largely unexplored.

The main aim of the present work is to provide a systematic thermodynamic study of this Curie--Weiss double-occupancy model and to determine its phase behavior over a wide range of the repulsion-to-attraction ratio. First, we show that the model can be obtained from the Blume--Capel model on a complete graph~\cite{HKBHK25} by a simple transformation. Then, we investigate its thermodynamic behavior within the grand canonical ensemble using the methods developed in~\cite{KKD20,KD22,RDKPS26a} for the cell model with unrestricted occupancy and Curie--Weiss interaction. Depending on the ratio between the repulsion and attraction couplings, the system exhibits qualitatively different types of phase behavior, including the regimes with one or two critical points, tricritical behavior, and triple points. For sufficiently strong repulsion, three fluid phases of different densities emerge, giving rise to both gas-liquid and liquid-liquid coexistence. We calculate the coordinates of the critical, tricritical, and triple points over a wide range of repulsion strengths and analyze the corresponding phase diagrams. The present model provides a minimal mean-field framework for investigating phase transitions and coexistence phenomena in multiple-occupancy lattice systems.

Before proceeding, we want to add a note about an existing ambiguity in the terminology of lattice gases. This ambiguity was noted by Stanley in his classic monograph~\cite[Appendix~A]{Stanley71}: ``{\it We say that a cell is in the occupied state if it is occupied by the centre of a molecule. It is worth emphasizing that we need not assume that the molecules themselves are restricted to cells, because some authors (erroneously) state that the atoms of the lattice-gas are restricted to the discrete sites of a lattice}''. Therefore, when considering fluid systems, it is more physically transparent to speak of the occupancy of cells by the centers of particles rather than regarding the particles as being restricted to a discrete set of lattice sites. To stress this, we call the studied model a ``cell'' model, rather than a lattice gas. It is worth noting, however, that the results associated with the occupancy-dependent configurational part of the partition function are the same in the two formulations. At the same time, retaining the momentum contribution enables one to recover the kinetic contribution required for the proper ideal-gas limit. This is especially important when calculating a quantity such as the entropy~\cite{RDKPS26b}.

	\section{Double occupancy cell fluid model with Curie-Weiss-type interaction}
We consider an open system of point particles in a macroscopic three-dimensional region ${\mathcal V} \subset \mathbb R^3$ of volume $V = \abs{\mathcal{V}}$. The region $\mathcal{V}$ is partitioned into $\mathscr{N}$ non-overlapping congruent cubic cells each of volume $v=\sigma^3$, where $\sigma$ is a linear length. The volume $V$ of the whole system is thus $V = v \mathscr{N}$. In the thermodynamic limit, $V \to \infty$, $\mathscr{N} \to \infty$, while $\lim_{V \to \infty; \mathscr{N} \to \infty}v = \lim_{V \to \infty; \mathscr{N} \to \infty} V/\mathscr{N} =\operatorname{const}$. The maximum number of particles allowed to occupy a cell is $n_{\textrm{max}} = 2$. Hence, the model is called the double-occupancy cell fluid model. In general, this model is isomorphic to the Blume-Capel model, and different interparticle interactions can be considered. In~\cite{LYZ21}, the case of nearest-neighbors attractions and in-cell repulsion was studied in two dimensions by means of Monte Carlo simulations. A case of a quantum system was studied in~\cite{Mancini05}, where the extended Hubbard model in the ionic limit was shown to be also isomorphic to the Blume-Capel model.

In the current study, we consider the interaction in the form of two contributions: a global, infinite-range attraction between each pair of particles, regardless of their position, and a local repulsion between two particles if they are located in the same cell. To begin with, we demonstrate how such a model appears from the Blume-Capel model on the complete graph~\cite{HKBHK25}. The Hamiltonian for the latter reads:
\begin{equation}
	\mathscr{H} = -\frac{J}{2\mathscr{N}} \sum_{l \neq m} S_l S_m + \Delta \sum_{l} S_l^2,
\end{equation}
where $J$ is the coupling constant, $\Delta$ is the so-called crystal field, $\mathscr{N}$ is the number of lattice sites, and the spin variables take on three values $S_l \in \{-1, 0, 1\}$, $l \in \{1, \ldots, \mathscr{N}\}$.

To transit to the cell-model representation, we introduce the cell-occupancy variables by
\begin{equation}
	n_l = S_l + 1, \quad n_l \in \{0,1,2\}.
\end{equation}
Performing the change of variable in the Hamiltonian, we arrive at
\begin{equation}
	\label{eq:hamiltonian}
	\mathscr{H} = -\frac{J}{2\mathscr{N}} \sum_{l \neq m} n_l n_m + \Delta \sum_l n_l (n_l - 1) + \left(J - \Delta - \frac{J}{\mathscr{N}}\right) \sum_l n_l,
\end{equation}
which clearly resembles the Hamiltonian considered in~\cite[equation~(1)]{LYZ21}.
The interpretation of different terms in the Hamiltonian is as follows. If two particles reside in different cells, they slightly attract with energy $-J/\mathscr{N} < 0$, which corresponds to the first term -- a global, Curie-Weiss-type attraction. When two particles are located in the same cell, there exist an energy cost $\Delta > 0$. This follows from the second term which does not equal zero only when a cell contains exactly two particles, $n_l=2$. In fact, the energy penalty for two particles being in the same cell is $2\Delta$, as follows from the second term in the Hamiltonian, since $n_l(n_l-1) =2$ for $n_l=2$. By contrast, in the Blume-Capel model the crystal-field term contributes only $\Delta$ for nonzero spin, since $S_l^2=1$ for $S_l = \pm 1$. Therefore, while $\Delta$ coincides with the crystal-field parameter of the Blume--Capel model under the mapping, the physical double-occupancy penalty is $2\Delta$. The last term in the Hamiltonian will just shift the chemical potential. Note that as a result of the variable change, an additive constant appeared in the Hamiltonian, which we have neglected.

In what follows, the standard notation for thermodynamic quantities is adopted: $T$ the absolute temperature; $\beta = (k_{\mathrm B} T)^{-1}$ the inverse temperature; $k_{\mathrm B}$ the Boltzmann constant; $P$ the pressure; $\mu$ the chemical potential; $N$ the number of particles; $\rho = N/V$ the particle number density; $m$ the particle mass; $\hbar$ the Planck constant. The natural units for physical quantities are as follows: $\sigma$ for the length, $v$ for the volume, $J$ for the energy.

As is also standard in the thermodynamics and statistical mechanics, we use the dimensionless, or reduced, quantities such as: the reduced temperature $T^* = k_{\mathrm B}T/J$; the reduced inverse temperature $\beta^* = \beta J$; the reduced pressure $P^* = Pv/J$; the reduced chemical potential $\mu^* = \mu/J$; the reduced particle number density $\rho^* = \rho \sigma^3$. It is worth noting that the definition of $\rho = N/V$, as well as other quantities expressed via $N$, is applicable for the ensembles with a fixed number of particles, e.g., for the canonical ensemble. In the case of ensembles with a variable (fluctuating) particle number, e.g., the grand canonical ensemble, this definition would read $\rho = \langle N \rangle / V$, where $\langle N \rangle$ is the average number of particles, and $\langle \ldots \rangle$ means the average over the corresponding ensemble. We will not introduce different notations for these two cases, as the meaning is clearly obvious depending on the considered ensemble.

Adding and subtracting the self-interaction term in the Hamiltonian~\eqref{eq:hamiltonian} via
\begin{equation}
	\sum_{l\neq m} n_l n_m = \sum_{l,m} n_l n_m - \sum_l n_l^2 = \left(\sum_l n_l\right)^2 - \sum_l n_l^2,
\end{equation}
we arrive at
\begin{equation}\label{NELA}
	\mathscr{H} = -\frac{J}{2\mathscr{N}} \left(\sum_l n_l\right)^2  + \left(\Delta + \frac{J}{2\mathscr{N}}\right) \sum_l n_l^2 + \left(J -2\Delta - \frac{J}{\mathscr{N}}\right) \sum_l n_l.
\end{equation}
Introducing the notation
\begin{equation}
	a = \frac{\Delta}{J},
\end{equation}
we can write the expression for the Boltzmann factor as:
\begin{equation}\label{BOF}
	{\textrm{e}}^{-\beta \mathscr{H}} = \exp \left[\frac{1}{2T^* \mathscr{N}} \left(\sum_l n_l\right)^2 + \frac{2a-1}{T^*} \sum_l n_l - \frac{a}{T^*} \sum_l n_l^2\right],
\end{equation}
where, in the thermodynamic limit $\mathscr{N}\to\infty$, we neglected contributions of the order $\mathscr{N}^{-1}$ appearing in the last two terms of \eqref{NELA}.

The grand partition function is now defined as:
\begin{equation}
	\label{eq:Xi_pi}
	\Xi  =  \sum_{n_1 = 0}^2 \left(\frac{v}{\Lambda^3}\right)^{n_1} \frac{{\textrm{e}}^{\beta\mu n_1}}{n_1!} \ldots \sum_{n_{\mathscr{N}} = 0}^2 \left(\frac{v}{\Lambda^3}\right)^{n_{\mathscr{N}}} \frac{{\textrm{e}}^{\beta\mu n_{\mathscr{N}}}}{n_{\mathscr{N}}!} {\textrm{e}}^{-\beta\mathscr{H}} ,
\end{equation}
which includes the Boltzmann factor from \eqref{BOF} and additional multipliers containing $\Lambda$, the de~Broglie thermal wavelength
\begin{equation}\label{Lam}
	\Lambda=\hbar\,\sqrt{\frac{2\piup\beta}m}
\end{equation}
with $\hbar$ being the reduced Planck constant and $m$ being the particle mass.

Usually, when lattice gas models are considered, the factors $(v/\Lambda^3)^{n_l}$ are not included, and only configuration partition functions are studied~\cite{FriedliVelenik17}. We retain these factors to maintain the proper thermodynamic connection to the underlying continuum system. Specifically, in \eqref{eq:Xi_pi}, the de~Broglie thermal wavelength $\Lambda$ appears due to the integration over momenta and taking into
account the factor $h^{-3}$ from the phase-space normalization in the definition of the grand partition function. Since we consider our cell model as a coarse-grained (discretized) version of a continuum system, the cell volume $v$ arises due to transition from the integration over space variables to the summation over lattice cells via
\begin{equation}
	\int_V {\textrm{d}} {\vb{r}} f(\vb{r}) = \sum_{l=1}^{\mathscr{N}} \int_{\Delta_l} {\textrm{d}} {\vb{r}} f(\vb{r}) \to v \sum_{l=1}^{\mathscr{N}}f_l \,.
\end{equation}
The issue of appearance of the factor $\frac{1}{\prod_l n_l!}$ in the definition of $\Xi$ was discussed in~\cite{FL18}. It was argued that its inclusion corresponds to the case of distinguishable particles, while its absence corresponds to the system of indistinguishable particles. The grand partition function for a system of indistinguishable particles is therefore expressed as
\begin{equation}
	\label{Xi:indist}
	\Xi =  \sum_{n_1 = 0}^2 \left(\frac{v}{\Lambda^3}\right)^{n_1} {\textrm{e}}^{\beta\mu n_1} \ldots \sum_{n_{\mathscr{N}} = 0}^2 \left(\frac{v}{\Lambda^3}\right)^{n_{\mathscr{N}}} {\textrm{e}}^{\beta\mu n_{\mathscr{N}}} {\textrm{e}}^{-\beta\mathscr{H}}.
\end{equation}
The comparison of the results for these two cases is of great interest. For example, in references~\cite{FL18, FL20} both cases were studied in parallel. In the current paper, we consider the case of the system of distinguishable particles more in detail, since it was the system we employed in our previous works. However, in section~\ref{sec:indist} we present the result for the system of indistinguishable particles as well. As will be seen there, the latter case can be more important regarding the isomorphism of the double-occupancy cell model to the Blume-Capel model.

	\section{Distinguishable particles}\label{sec:distinct}

In this Section, we focus on the case of distinguishable particles.
Using the explicit expression for the Boltzmann factor \eqref{BOF} in the grand partition function~\eqref{eq:Xi_pi} we obtain
\begin{eqnarray}
	\label{eq:Xi1}
	\Xi & = & \sum_{n_1 = 0}^2 \cdots \sum_{n_{\mathscr{N}} = 0}^2 \exp\left[\frac{1}{2T^*\mathscr{N}} \left(\sum_l n_l\right)^2\right] \frac{\left(v^* T^{*3/2}\right)^{\sum_l n_l}}{n_1! \ldots n_{\mathscr{N}}!} \exp\left[\frac{\mu^* + 2a - 1}{T^*}\sum_l n_l - \frac{a}{T^*}\sum_l n_l^2\right]
	\nonumber\\
	& = & \sum_{n_1 = 0}^2 \cdots \sum_{n_{\mathscr{N}} = 0}^2 \exp\left[\frac{1}{2T^*\mathscr{N}} \left(\sum_l n_l\right)^2\right] \prod_l \pi(T^*,\mu^*;n_l),
\end{eqnarray}
where
\begin{equation}\label{eq:Xi1X}
	\pi(T^*,\mu^*; n_l) = \frac{\left(v^* T^{*3/2}\right)^{n_l}}{n_l!} \exp\left(\frac{\mu^* + 2a -1}{T^*}n_l - \frac{a}{T^*} n_l^2\right).
\end{equation}
In \eqref{eq:Xi1} and \eqref{eq:Xi1X}, we introduced the dimensionless cell volume $v^*$ via $v^* = v/\Lambda_J^3$, where $\Lambda_J\equiv(2\piup \hbar^2/mJ)^{1/2}$ stems from the definition of the de~Broglie thermal wavelength in \eqref{Lam}.
Equations \eqref{eq:Xi1} and \eqref{eq:Xi1X} are of the same functional form as in~\cite[(2.11)]{KKD20} for the unrestricted occupancy cell model, with three differences: now, the summation over occupancy numbers is restricted by the maximum occupancy number $n_{\textrm{max}}=2$; the chemical potential is shifted by value $2a-1$; the coefficient at $n_l^2$ is $a/T^*$ instead of $a/(2T^*)$ in~\cite[(2.11)]{KKD20}. This allows the application of the methodology developed in~\cite{KKD20,KD22,RDKPS26a} to the present case.

Our transformation of the Blume-Capel model on a complete graph to a double-occupancy cell fluid model with the Curie-Weiss interaction establishes an equivalence in mathematical description of these two models.
There exists one more crucial difference between the unrestricted- and double-occupancy models with Curie-Weiss-type interaction. The stability condition $a > 1$~\cite{KKD20} is not required anymore, since the system's collapse~\cite{FR66,HR07,FL20} is easily avoided by a finite maximum occupancy number $n_{\textrm{max}}=2$. This allows us to consider the system in a wider range of parameter $a$. In fact, there are no restrictions on $a$ in the double-occupancy model, and it can take on any real value. In this paper, we consider only non-negative values $a \geqslant 0$.

Let us proceed with the evaluation of the grand partition function $\Xi$ from \eqref{eq:Xi1}, following the methodology of~\cite{KKD20,KD22,RDKPS26a}. The main steps consist in linearizing the quadratic Curie-Weiss interaction using a Gaussian integral, evaluating the resulting integral in the thermodynamic limit by Laplace's method, and obtaining the thermodynamic quantities from the global maximum of its exponent.
In order to explicitly evaluate the repeated sums over $n_1,\ldots,n_{\mathscr{N}}$, we have to linearize the sum $\sum_l n_l$ in the exponential. This is achieved by representing the relevant factor through the Gaussian integral, viz.
\begin{equation}
	\exp\left[\frac{1}{2T^*\mathscr{N}} \left(\sum_l n_l\right)^2\right] = \sqrt{\frac{\mathscr{N}T^*}{2\piup}} \int_{-\infty}^{\infty} \exp\left(-\frac{\mathscr{N}}{2}T^* y^2 + y \sum_l n_l\right) {\textrm{d}} y.
\end{equation}
To simplify the notation, we introduce the effective (shifted) chemical potential $\mu_{\textrm{eff}}^*$ via
\begin{equation}
	\mu_{\textrm{eff}}^* = \mu^* + 2a -1.
\end{equation}
As a result, we obtain the grand partition function $\Xi$ in the form
\begin{equation}
	\label{Y1}
	\Xi(T^*,\mu_{\textrm{eff}}^*) = \sqrt{\frac{\mathscr{N} T^*}{2\piup}} \int\limits_{-\infty}^{\infty} \exp\left[\mathscr{N} \tilde{E}(T^*,\mu_{\textrm{eff}}^*; y)\right] {\textrm{d}} y
\end{equation}
with
\begin{equation}
	\tilde{E}(T^*,\mu_{\textrm{eff}}^*; y) = -\frac{1}{2}T^* y^2 + \ln \tilde{K}_0(T^*,\mu_{\textrm{eff}}^*;y)\, .
\end{equation}
Due to summations over $n_1,\ldots,n_{\mathscr{N}}$ in \eqref{eq:Xi1}, the  special function $\tilde{K}_0(T^*,\mu_{\textrm{eff}}^*;y)$ appears in $\tilde{E}(T^*,\mu_{\textrm{eff}}^*; y)$, which is defined as
\begin{equation}
	\label{Y3}
	\tilde{K}_0(T^*,\mu_{\textrm{eff}}^*; y) = \sum_{n=0}^{2} \frac{\left(v^* T^{*3/2}\right)^n}{n!} \exp \left[\left(y + \frac{\mu_{\textrm{eff}}^*}{T^*}\right)n - \frac{a}{T^*}n^2\right],
\end{equation}
and represents the partition sum of a single cell.

Following~\cite{KD22}, we shift the integration variable $y$ in \eqref{Y1}--\eqref{Y3} via
\begin{equation}
	z = y + \frac{\mu^*_{\textrm{eff}}}{T^*} + \ln v^*
\end{equation}
thereby eliminating the dependencies on chemical potential and cell volume from the function $\tilde{K}_0$. Thus, we arrive at the integral representation
\begin{equation}
	\label{eq:Xi_z}
	\Xi = \sqrt{\frac{\mathscr{N} T^*}{2\piup}} \int\limits_{-\infty}^{\infty} \exp \left[\mathscr{N} E(T^*, \mu^*_{\textrm{eff}}; z)\right] {\textrm{d}} z,
\end{equation}
with
\begin{equation}
	\label{def:E}
	E(T^*,\mu^*;z) = -\frac{1}{2}T^* \left(z - \frac{\mu^*_{\textrm{eff}}}{T^*} - \ln v^*\right)^2 + \ln K_0(T^*;z),
\end{equation}
and
\begin{equation}
	\label{def:K0}
	K_0(T^*;z) = \sum_{n=0}^{2} \frac{\left(T^{*3/2}\right)^n}{n!} \exp\left(zn - \frac{a}{T^*}n^2\right).
\end{equation}
To evaluate the integral in equation~\eqref{eq:Xi_z}, we apply Laplace's method (see, e.g.,~\cite{BenderOrszag99}). In the large-$\mathscr{N}$ limit, physically  corresponding to the thermodynamic limit, the integral is dominated by the global maximum of the exponent, yielding
\begin{equation}\label{Laplace}
	\Xi \simeq \exp\left[\mathscr{N} E(T^*,\mu^*_{\textrm{eff}}; \bar{z}_{\textrm{max}})\right],
\end{equation}
where $\bar{z}_{\textrm{max}} = \bar{z}_{\textrm{max}}(T^*,\mu^*_{\textrm{eff}})$ is the coordinate of the global maximum of $E(T^*,\mu^*_{\textrm{eff}}; z)$.
The location of this maximum is determined by the extremum condition (cf.~\cite{KKD20,KD22,RDKPS26a})
\begin{equation}
	\label{eq:extremum}
	T^* \bar{z} - \left(\mu^*_{\textrm{eff}} + T^* \ln v^*\right) = \frac{K_1(T^*; \bar{z})}{K_0(T^*; \bar{z})},
\end{equation}
where $K_1(T^*; z)$ is the first member of the hierarchy of functions $K_j(T^*; z)$, $j=1, 2, \ldots $, with 
\begin{equation}
	\label{def:Kj}
	K_j(T^*; z) = \sum_{n=0}^2 \frac{n^j \left(T^{*3/2}\right)^n}{n!} \exp\left(z n - \frac{a}{T^*}n^2\right)
\end{equation}
 being the $j$'th derivatives of $K_0(T^*;z)$ from \eqref{def:K0} with respect to $z$.

Solving the equation~\eqref{eq:extremum}, we select only the solutions corresponding to global maxima of $E(T^*, \mu^*_{\textrm{eff}}; z)$ and denote them by $\bar{z}_{\textrm{max}}$. A useful consequence from equation~\eqref{eq:extremum} is the relation
\begin{equation}
	\label{eq:extremum_z}
	\bar{z}_{\textrm{max}} - \frac{\mu^*_{\textrm{eff}}}{T^*} - \ln v^* = \frac{1}{T^*} \frac{K_1(T^*; \bar{z}_{\textrm{max}})}{K_0(T^*; \bar{z}_{\textrm{max}})}.
\end{equation}
Substituting this back into the function $E$ eliminates its explicit dependence on $\mu^*_{\textrm{eff}}$, and we arrive at
\begin{equation}\label{Esimplified}
	E(T^*; \bar{z}_{\textrm{max}}) = -\frac{1}{2T^*} \left[\frac{K_1(T^*; \bar{z}_{\textrm{max}})}{K_0(T^*; \bar{z}_{\textrm{max}})}\right]^2 + \ln K_0(T^*;\bar{z}_{\textrm{max}}).
\end{equation}

By standard formulas of statistical mechanics, we obtain the pressure as a function of the temperature (explicitly) and the chemical potential (implicitly) via
\begin{equation}
	\label{eq:pressure}
	P^* = T^* E(T^*; \bar{z}_{\textrm{max}}).
\end{equation}
The reduced particle number density (or the average occupancy number) is then given by
\begin{equation}
	\label{eq:rho}
	\rho^* = \frac{K_1(T^*; \bar{z}_{\textrm{max}})}{K_0(T^*; \bar{z}_{\textrm{max}})}.
\end{equation}
At any given temperature $T^*$, equations~\eqref{eq:pressure} and~\eqref{eq:rho} constitute a parametric equation of state with the parameter $\bar{z}_{\textrm{max}}$ for the double-occupancy cell model with Curie-Weiss interaction.

	\subsection{Critical point coordinates}
	\label{sec:cp}
	
	\begin{table}[h]
		\centering
		\caption{Critical point coordinates of the double-occupancy cell model for the case of distinguishable particles. At the tricritical point, $a_{\textrm{tc}} = \frac{1}{2} \ln 2$. For any value $a \leqslant a_{\textrm{tc}}$, there is a single critical point; for any value $a > a_{\textrm{tc}}$, there are two critical points. The large-$a$ critical pressure $P^{*(1)}_c=\frac{1}{4}\left(\ln 2 - \frac{1}{2}\right)$ has been analytically obtained in \cite[section~9.4]{DKPRS26arxiv}.}
		\begin{tabular}{c|c|c|c|c|c}
			\toprule
			$a=\Delta/J$ & $T^*_c$ & $\rho^{*(1)}_c $ & $P^{*(1)}_c$ & $\rho^{*(2)}_c$ & $P^{*(2)}_c$\\
			\midrule
			5.0 & $0.25 = \frac{1}{4}$ & $0.5 = \frac{1}{2}$ & $0.0482868 \approx \frac{1}{4}\left(\ln 2 - \frac{1}{2}\right)$ & $1.5 = \frac{3}{2}$ & 9.22157 \\
			2.0 & 0.250000 & 0.500000 & 0.0482868 & 1.50000 & 3.22157 \\
			1.5 & 0.250003 & 0.500009 & 0.0482882 & 1.49999 & 2.22158 \\
			1.2 & 0.250034 & 0.500102 & 0.0483018 & 1.49990 & 1.62166 \\
			1.0 & 0.250168 & 0.500507 & 0.0483616 & 1.49949 & 1.22202 \\
			0.9 & 0.250378 & 0.501135 & 0.0484545 & 1.49886 & 1.02257 \\
			0.8 & 0.250852 & 0.502562 & 0.0486653 & 1.49744 & 0.823816 \\
			0.7 & 0.251946 & 0.505871 & 0.0491542 & 1.49413 & 0.626683 \\
			0.6 & 0.254567 & 0.513894 & 0.0503395 & 1.48611 & 0.433501 \\
			0.5 & 0.261400 & 0.535459 & 0.0535217 & 1.46454 & 0.250934 \\
			0.4 & 0.283929 & 0.615014 & 0.0650251 & 1.38499 & 0.104907 \\
			$\frac{1}{2}\ln 2 \approx 0.346574$ & $\frac{1}{3} \approx 0.333333$ & 1.00000 & 0.0972532 & 1.00000 & 0.0972532\\
			0.3 & 0.400889 & 1.0 & 0.144316 & - & - \\
			0.2 & 0.483166 & 1.0 & 0.186358 & - & - \\
			$\frac{1}{4} \ln 2 \approx 0.173287$ & $\frac{1}{2}$ & 1.0 & 0.193147 & - & - \\
			0.1 & 0.540284 & 1.0 & 0.207128 & - & - \\
			0.0 & 0.585786 & 1.0 & 0.219315 & - & - \\
			\bottomrule
		\end{tabular}
		\label{tab:critical-points}
	\end{table}
	As was shown in~\cite{RDKPS26a}, the physical conditions for finding the critical point
	\begin{equation}
		\left(\frac{\partial P^*}{\partial \rho^*}\right)_{T^*} = 0 \quad {\text {and} } \quad
		\left(\frac{\partial^2 P^*}{\partial \rho^*}\right)_{T^*} = 0,
	\end{equation}
	in combination with the extremum condition~\eqref{eq:extremum} for $E(T^*, \mu^*_{\textrm{eff}}; z)$, result into the system of equations
	\begin{equation}
		\label{eq:system}
		\begin{cases}
			E_1(T^*,\mu^*_{\textrm{eff}}; \bar{z}) = 0,\\
			E_2(T^*; \bar{z}) = 0,\\
			E_3(T^*; \bar{z}) = 0.
		\end{cases}
	\end{equation}
	where functions $E_j$ are defined as derivatives of the order $j$ of $E(T^*,\mu^*_{\textrm{eff}}; z)$ with respect to $z$:
	\begin{equation}
		E_j(T^*,\mu^*_{\textrm{eff}}; z) \equiv \frac{\partial^j}{\partial z^j} E(T^*, \mu^*_{\textrm{eff}}; z).
	\end{equation}
	Note that, since the function $E(T^*,\mu^*_{\textrm{eff}};z)$ in~\eqref{def:E} contains at most the second power of $\mu^*_{\textrm{eff}}$, its derivatives $E_j$ with $j \geqslant 2$ do not explicitly depend on $\mu^*_{\textrm{eff}}$. Therefore, the last two equations in \eqref{eq:system} directly determine the critical values for the temperature, $T^*_c$, and for $\bar{z}^{(c)}_{\textrm{max}}$. Then, we find the critical value of the chemical potential $\mu^{*(c)}_{\textrm{eff}}$ from the first condition $E_1 = 0$, which is equivalent to equation~\eqref{eq:extremum_z}:
	\begin{equation} \label{eq:mueff}
		\mu^*_{\textrm{eff}} = T^* \bar{z}_{\textrm{max}} - \frac{K_1(T^*; \bar{z}_{\textrm{max}})}{K_0(T^*; \bar{z}_{\textrm{max}})} - T^* \ln v^*.
	\end{equation}
	Subsequently, we determine all other quantities at the critical point, such as pressure via~\eqref{eq:pressure} or the particle number density via~\eqref{eq:rho}.
	
\begin{figure}
		\centering
		\includegraphics[width=0.5\textwidth]{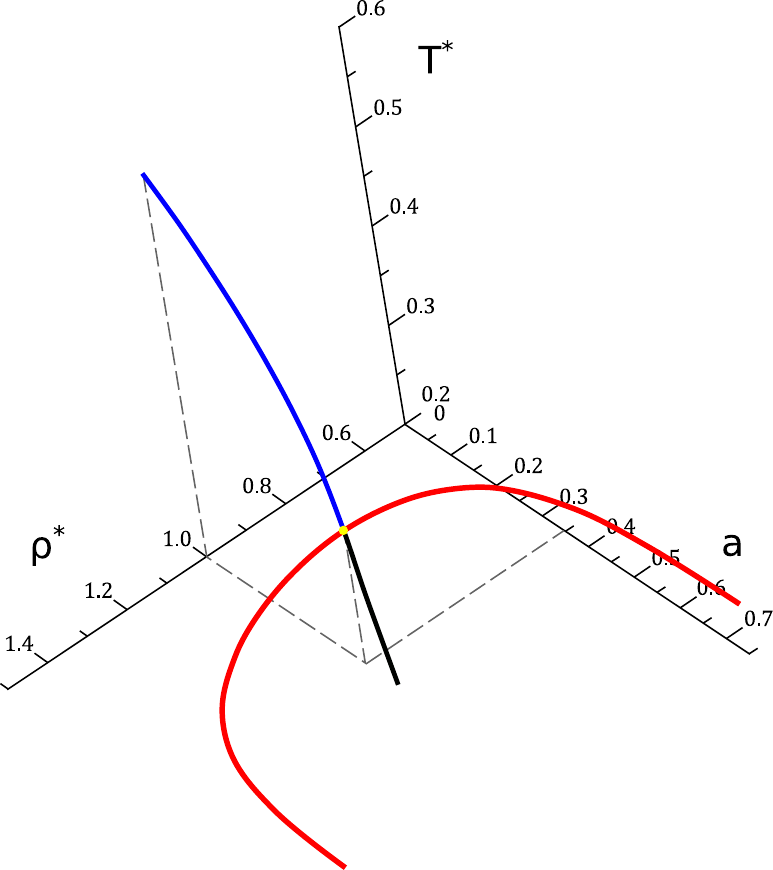}
		\caption{(Colour online) Phase diagram of the double-occupancy cell model in the case of distinguishable particles. Blue curve: the line of critical points at $a < a_{\textrm{tc}}$. Red curves: the lines of critical points at $a > a_{\textrm{tc}}$. Black curve: the line of triple points in the interval $a_{\textrm{tc}} < a < 0.5$. All the lines meet at the tricritical point with $a=a_{\textrm{tc}}=\frac{1}{2} \ln 2$.}
		\label{fig:T-rho-a-3d}
	\end{figure}

	It turns out that the number of solutions to the system of equations~\eqref{eq:system} depends on the ratio $a=\Delta/J$, and therefore the phase diagram also strongly depends  on the value of this parameter. For large values of $a$, there are two critical points at the same critical temperature but at different critical values of $\bar{z}_{\textrm{max}}$, which in turn means two different critical densities by equation~\eqref{eq:rho}. For example, for $a=5.0$, the two critical points $(T^*_c; \rho^*_c)$ have coordinates $(0.25; 0.5)$ and $(0.25;1.5)$. These values of the critical point coordinates are limiting at large $a$, and can be written as $(1/4; 1/2)$ and $(1/4; 3/2)$, respectively. As the value of $a$ decreases, the critical temperature increases, the lower critical density increases above $0.5$, while the higher critical density decreases below $1.5$. For example, at $a=0.6$, the two critical points have the coordinates $(0.254567; 0.513894)$ and $(0.254567; 1.48611)$, correspondingly. More data are collected in table~\ref{tab:critical-points}. We denote the quantities related to the lower-density critical point by a superscript~${}^{(1)}$, and to the higher-density critical point by ${}^{(2)}$. We observe that as $a$ decreases, the difference between $\rho^{*(1)}_c$ and $\rho^{*(2)}_c$ becomes smaller and smaller, until they eventually become the same at $a_{\textrm{tc}}=\frac{1}{2}\ln 2 \approx 0.346574$, where the two lines of critical points merge into one at the point $(0.333333; 1.00000)$, or simply $(1/3;1)$. This case corresponds to a tricritical point. For $a < a_{\textrm{tc}}$, only one critical point remains in the phase diagram, always with the critical density $\rho^*_c = 1.0$. As the parameter $a$ changes from $a_{\textrm{tc}}$ down to 0, the critical temperature rises from $T^*_t = 0.333333$ to $T^*_c = 0.585786$. Figure~\ref{fig:T-rho-a-3d} illustrates this behavior based on the data from table~\ref{tab:critical-points}.
	
	In the current study, due to long-range nature of the Curie-Weiss interaction, all the critical points are of the mean-field kind. However, in the case of short range interactions, e.g., nearest-neighbours attraction and in-cell repulsion, we expect that the critical points would belong to the three-dimensional Ising universality class, as they do in the Blume-Capel model~\cite{JL80, GKRT01}. This topic requires a separate investigation.
	
	Note that in the region where two critical points exist, the sum of two critical densities always remains the same:
	\begin{equation}
		\rho^{*(1)}_c + \rho^{*(2)}_c = 2.
	\end{equation}
	
	\subsection{Coexistence curves}\label{sec:coex}
	Having found critical point coordinates for different values of $a$ in section~\ref{sec:cp}, we now construct the coexistence curves. 
	The coexistence states are determined from the physical conditions that the pressure and chemical potential are equal in two phases at a given temperature. Using equations~\eqref{eq:pressure} and~\eqref{eq:mueff}, these conditions can be written, for example for the coexistence between phases I and II, as:
	\begin{equation}\label{eq:coex}
		\left\{
		\begin{aligned}
			E(T^*; z_{\rm I}) & = E(T^*; z_{\rm II}),
			\\
			\mu^*_{\textrm{eff}}(T^*; z_{\rm I}) & = \mu^*_{\textrm{eff}}(T^*; z_{\rm II}).
		\end{aligned}
		\right.
	\end{equation}
	Here, $z_{\rm I}$ and $z_{\rm II}$ correspond to two distinct global maxima of the function $E$, associated with phases I and II, respectively. Below the corresponding critical temperature, equation~\eqref{eq:coex} has a pair of solutions ($z_{\rm I}, z_{\rm II}$), one on each of the two phase branches. Substitution of these values into equation~\eqref{eq:rho} gives the two coexisting densities, $\rho^*_{\rm I}$ and $\rho^*_{\rm II}$. Repeating this procedure at different temperatures generates the corresponding coexistence curve in the $(T^*-\rho^*)$ plane. The same procedure is applied to the coexistence between phases II and III.
	
	\begin{figure}
		\centering
		\begin{subfigure}[b]{0.23\textwidth}
			\centering
			\includegraphics[width=\textwidth]{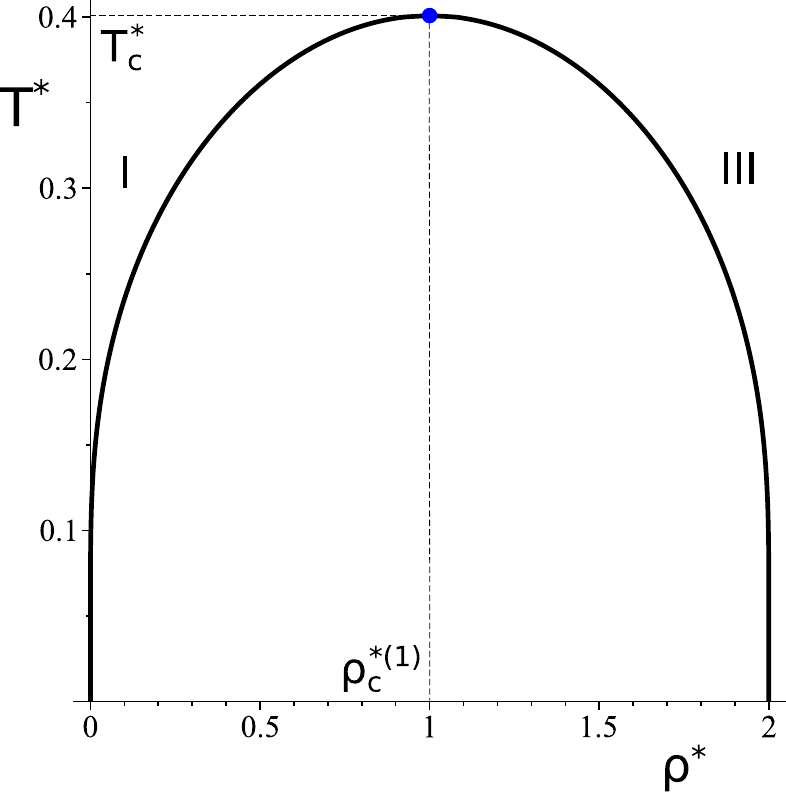}
			\caption{$a=0.3$}
			\label{fig:pd1}
		\end{subfigure}
		\hfill
		\begin{subfigure}[b]{0.23\textwidth}
			\centering
			\includegraphics[width=\textwidth]{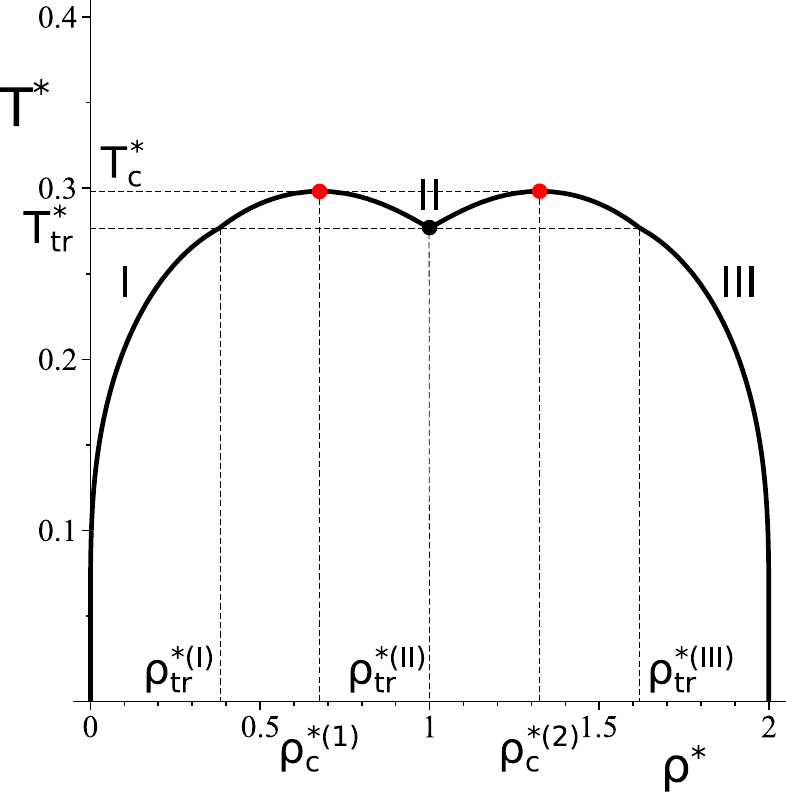}
			\caption{$a=0.375$}
			\label{fig:pd2}
		\end{subfigure}
		\hfill
		\begin{subfigure}[b]{0.23\textwidth}
			\centering
			\includegraphics[width=\textwidth]{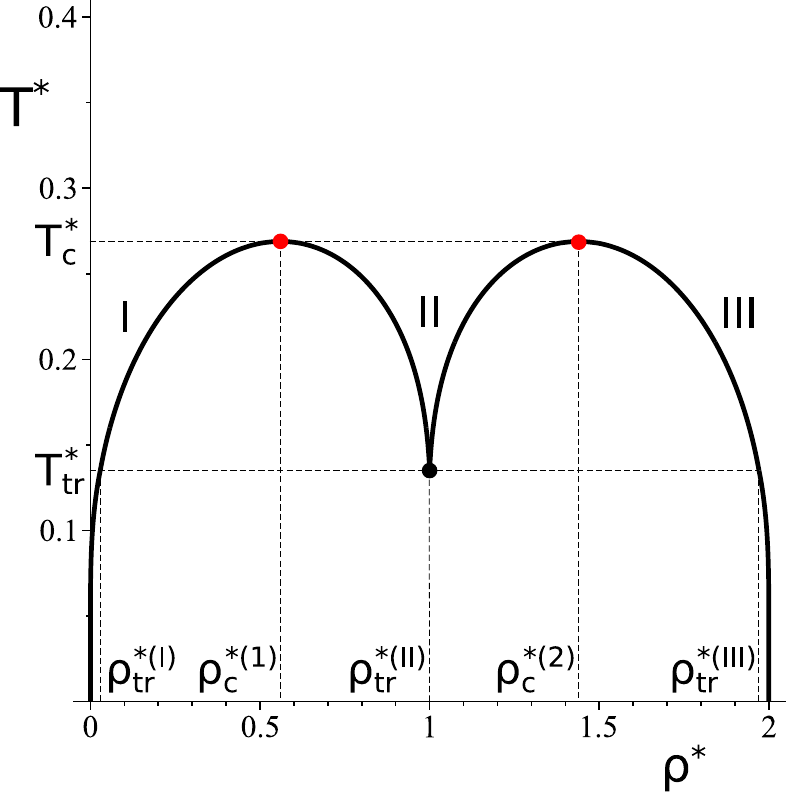}
			\caption{$a=0.45$}
			\label{fig:pd3}
		\end{subfigure}
		\hfill
		\begin{subfigure}[b]{0.23\textwidth}
			\centering
			\includegraphics[width=\textwidth]{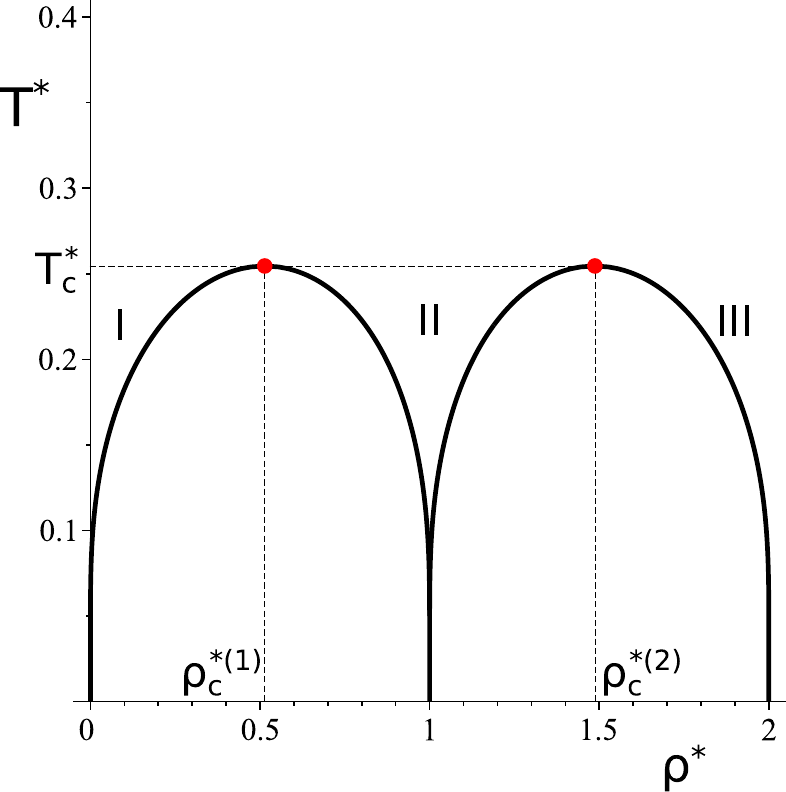}
			\caption{$a=0.6$}
			\label{fig:pd4}
		\end{subfigure}
		\caption{(Colour online) The $T^* - \rho^*$ phase diagrams of the double-occupancy cell model in the case of distinguishable particles at different values of $a=\Delta/J$. Solid curves are the coexistence lines. {\textbf{(a)}}:~The coexistence line between phases I and III ends at the critical point. {\textbf{(b)}} and {\textbf{(c)}}: Coexistence lines between phases I --- II and II --- III are present in the temperature interval $T^*_{\textrm{tr}} \leqslant T^* \leqslant T^*_c$, each ending at its own critical point. For $T^* \leqslant T^*_{\textrm{tr}}$, phases I and III coexist. $T^*_{\textrm{tr}}$ is the temperature of the triple point. {\textbf{(d)}}:~Coexistence lines between phases I --- II and II --- III are present for $0 \leqslant T^* \leqslant T^*_c$, each ending at its own critical point.}
		\label{fig:phase_diagrams}
		
	\end{figure}
	
	The $T^* - \rho^*$ phase diagrams are shown in figure~\ref{fig:phase_diagrams}, for $a=0.3,$ $0.375,$ $0.45,$ and $0.6$.
	At small $a$ (weak repulsion), the phase behavior of the double-occupancy cell model resembles the phase behavior of a single-occupancy lattice gas: there is a single critical point, and at lower temperatures there is a symmetric curve of two-phase coexistence. Figure~\ref{fig:pd1} shows the $T^* - \rho^*$ phase diagram at $a=0.3$, typical of small values of $a$ in the range $a \leqslant a_{\textrm{tc}}= \frac{1}{2}\,\ln 2$. At $a=0.3$, the coordinates of the critical point are $T^*_c = 0.401$, $\rho^*_c=1.0$, and $P^*_c=0.144$.
	
	Once $a$ exceeds $\frac{1}{2}\ln 2$, the coexistence line bifurcates, and two critical points appear. Figure~\ref{fig:pd2} illustrates the phase diagram at $a=0.375$, where two critical points exist at the same critical temperature $T^*_c = 0.298$, but with different densities, $\rho^{*(1)}_c = 0.676$ and $\rho^{*(2)}_c=1.32$, and different pressures: $P^{*(1)}_c=0.073$ and $P^{*(2)}_c=0.086$.
	In this regime, an intermediate-density phase (phase II) emerges  between the low-density phase I and high-density phase III. However, the phase II exists only above some temperature $T^*_{\textrm{tr}}$. Below it, the phase transition occurs directly between phases I and III. The point $(T^*_{\textrm{tr}}, P^*_{\textrm{tr}})$ is the \textit{triple point}. Here, all three phases coexist at the same pressure and temperature, but at three different densities.
	
	As $a$ further rises, the critical temperature drops slowly, while the difference between the critical densities increases, and the triple-point temperature decreases quite notably. Figures~\ref{fig:pd2} ($a=0.375$) and~\ref{fig:pd3} ($a=0.45$) illustrate this progress in phase behavior.
	
	At $a \approx 0.5$, the triple-point temperature becomes zero, and no triple-point exists for greater $a$. This situation is illustrated in figure~\ref{fig:pd4} with $a=0.6$. At such high values of $a$, three distinct phases exist at all low temperatures. When temperatures are very low, e.g., $T^* < 0.05$, the phase densities remain nearly constant with $\rho^*_{I} \approx 0$, $\rho^*_{II} \approx 1$, and $\rho^*_{III} \approx 2$. This reveals pronounced phase transitions between three phases with integer-valued densities.
	
\begin{figure}
		\centering
		\includegraphics[width=0.5\textwidth]{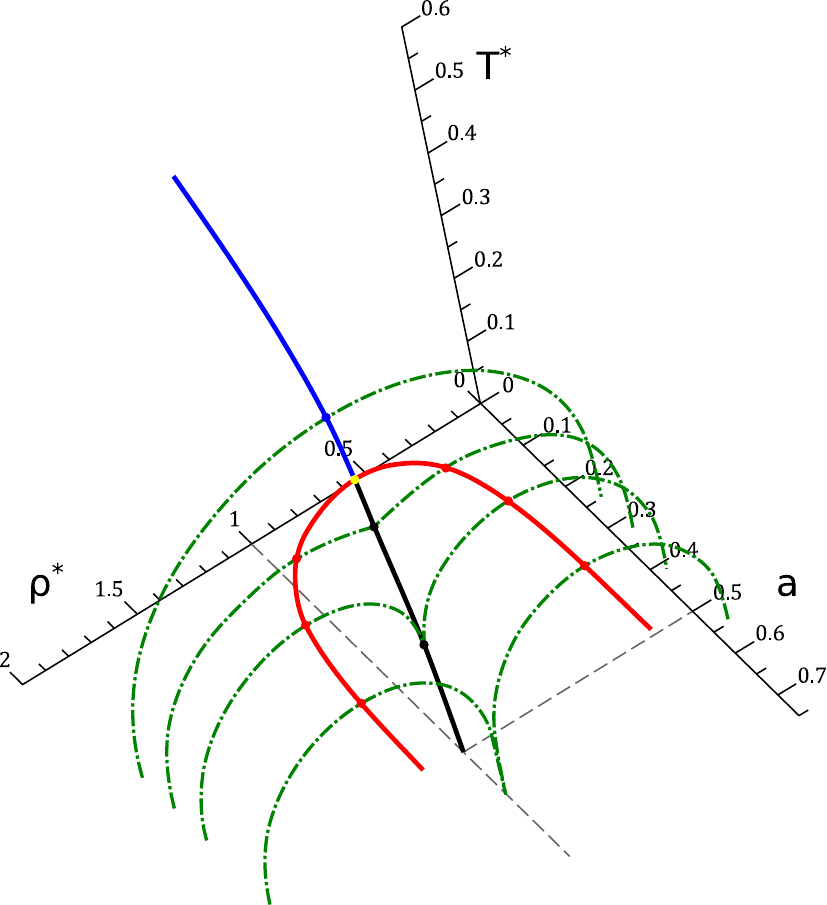}
		\caption{(Colour online) The phase diagram in the $(T^*,\rho^*,a)$-space for the case of distinguishable particles. The figure combines representations of figures~\ref{fig:T-rho-a-3d} and~\ref{fig:phase_diagrams} in one view. Green dashed curves: the coexistence lines at $a=0.3$, $a=0.375$, $a=0.45$, and $a=0.6$. Blue and red curves are the critical-point lines. Black curve is the triple-point line.}
		\label{fig:combined}
	\end{figure}

	Figure~\ref{fig:combined}  combines the results presented in figures~\ref{fig:T-rho-a-3d} and~\ref{fig:phase_diagrams}. The coexistence curves are presented in the three dimensional space along with the critical point lines and the triple-point line.
	
	\subsection{Triple point coordinates}
	The triple point is determined in a similar way. In this case, at a certain temperature $T^*_{\rm tr}$ and chemical potential $\mu^*_{\textrm{eff},{\rm tr}}$, the function $E(T^*_{\rm tr}; z)$ possesses three distinct global maxima, $z_{\rm I}$, $z_{\rm II}$, and $z_{\rm III}$. At the triple point, the corresponding three phases have the same pressure and chemical potential and can therefore coexist.
	Mathematically, this is expressed by the following system of equations:
	\begin{equation}
		\left\{
		\begin{aligned}
			E(T^*_{\rm tr}; z_{\rm I}) & = & E(T^*_{\rm tr}; z_{\rm II}) & = & E(T^*_{\rm tr}; z_{\rm III}),
			\\
			\mu^*_{\textrm{eff}}(T^*_{\rm tr}; z_{\rm I}) & = & \mu^*_{\textrm{eff}}(T^*_{\rm tr}; z_{\rm II}) & = & \mu^*_{\textrm{eff}}(T^*_{\rm tr}; z_{\rm III}).
		\end{aligned}
		\right.
	\end{equation}
	The corresponding densities $\rho^*_{\rm I}$, $\rho^*_{\rm II}$, and $\rho^*_{\rm III}$ are obtained from equation~\eqref{eq:rho} by evaluating it at the three respective values of $z$. The triple-point coordinates reported below are obtained numerically from these conditions.
	
	For each value of $a$ in the interval $\frac{1}{2}\ln 2 < a < 0.5$, the system possesses a triple point. At this point, three phases with different densities --- I, II, and III --- can coexist at the same temperature and pressure. The numerical results for triple-point coordinates are presented in table~\ref{tab:triple-line}.

	The triple-point temperature decreases more rapidly with increasing $a$ than the critical temperature. Consequently, $T^*_{\textrm{tr}} \leqslant T^*_{c}$ for every $a > a_{\textrm{tc}}$, with equality only at the tricritical point. As $a$ approaches 0.5, the triple-point temperature tends to zero, and for $a\geqslant 0.5$, the triple point no longer exists. Similarly, the triple-point pressure decreases from its tricritical value $P^*_{\textrm{tc}} = 0.097$ to $P^*_{\textrm{tr}} = 0$ at $a=0.5$. The evolution of the triple-point densities differs for the three coexisting phases. At the tricritical point, they are the same for three phases $\rho^*_{\textrm{tc}} = 1.0$. As $a$ increases above $a_{\textrm{tc}}$, the triple-point density of phase I gradually decreases from $\rho^{*(I)}_{\textrm{tr}} = 1.0$ to $\rho^{*(I)}_{\textrm{tr}} = 0.0$, while the triple-point density of phase III simultaneously increases from $\rho^{*(III)}_{\textrm{tr}} = 1.0$ to $\rho^{*(III)}_{\textrm{tr}} = 2.0$. Throughout this interval, the triple-point density of phase II remains constant at $\rho^{*(II)}_{\textrm{tr}} = 1.0$. Thus, the tricritical point marks the onset of the triple-point line. As the repulsion strength increases further, the densities of the three coexisting phases gradually approach their limiting low-temperature values 0, 1, and 2, characteristic of phases I, II, and III, respectively.
	
	\begin{table}[h]
		\centering
		\caption{Triple-point coordinates of the double-occupancy cell model with distinguishable particles. Triple points exist in the interval $a_{\textrm{tc}} < \Delta/J < 0.5$, where $a_{\textrm{tc}} = \frac{1}{2} \ln 2$ is the value of $a$ at the \emph{tricritical} point.}
		\begin{tabular}{c|c|c|c|c|c}
			\toprule
			$a=\Delta/J$ & $T^*_{\textrm{tr}}$ & $P^*_{\textrm{tr}}$ & $\rho^{*(I)}_{\textrm{tr}}$ & $\rho^{*(II)}_{\textrm{tr}}$ & $\rho^{*(III)}_{\textrm{tr}}$ \\
			\midrule
			0.500 & $\approx 0.0$ & $\approx 0.0$ & $\approx 0.0$ & 1.0 & $\approx 2.0$ \\
			0.475 & 0.0719370 & $6.93 \cdot 10^{-5}$ & $9.69 \cdot 10^{-4}$  & 1.0 & 1.99903 \\
			0.450 & 0.135275 & $3.585 \cdot 10^{-3}$ & 0.0292224 & 1.0 & 1.97078 \\
			0.425 & 0.185875 & 0.0143790 & 0.0981956 & 1.0 & 1.90180 \\
			0.400 & 0.231382 & 0.0320139 & 0.207212 & 1.0 & 1.79279 \\
			0.375 & 0.276725 & 0.0569574 & 0.380556 & 1.0 & 1.61944 \\
			0.350 & 0.325987 & 0.0915080 & 0.766700 & 1.0 & 1.23330 \\
			$\frac{1}{2}\ln 2 \approx 0.346574$ & $0.333333 = \frac{1}{3}$ & 0.0972532 & 1.0 & 1.0 & 1.0 \\
			\bottomrule
		\end{tabular}
		\label{tab:triple-line}
	\end{table}
	
	It is well known that in the classical single-occupancy lattice gas models only two phases are possible, and thus triple points do not occur. However, one or more triple points can exist in multiple-occupancy models. In our previous studies of the cell model with unrestricted cell occupancy, which resulted in a strict limitation $a \geqslant 1$, the triple points could not be observed~\cite{KKD20,KD22,DKPP26}.
	Only recently~\cite{KDPRS26}, we managed to obtain a triple point in this model, by allowing the interparticle interactions to depend on temperature, and thus become phase-dependent.
	In the double-occupancy model, there is no restriction $a \geqslant 1$ on $a$, and triple points naturally emerge in the interval $a_{\textrm{tc}} < a < 0.5$, not accessible before~\cite{KKD20,KD22,DKPP26}.
	
	\subsection{Possible issues and their resolution}
	Let us compare some results from tables~\ref{tab:critical-points} and \ref{tab:triple-line} with their counterparts in reference~\cite{HKBHK25}. At $\Delta = 0$, the critical temperature of the double-occupancy cell model is $T^*_c=0.586$, while for the Blume-Capel model considered in~\cite{HKBHK25} it was found to be $T^*_c=2/3$. Moreover, at the tricritical point $a_{\textrm{tc}}=\frac{1}{2}\ln 2$ in the cell model, while $a_{\textrm{tc}}=\frac{2}{3} \ln 2$ in its Blume-Capel counterpart. We find out that these discrepancies are associated with the particle  distinguishability assumed in the current section, which manifests itself in  multipliers $\frac{1}{n_l!}$ in the definition of the grand partition function.
	
	\section{Indistinguishable particles} \label{sec:indist}
	For a system of indistinguishable particles, the grand partition function is given by~\eqref{Xi:indist}. The calculation goes in parallel with that in section~\ref{sec:distinct}, and the results~\eqref{eq:Xi_z}--\eqref{eq:rho} remain the same, but with modified special functions
	\begin{equation}\label{def:K0ind}
		K_j(T^*; z) = \sum_{n=0}^2 n^j \left(T^{*3/2}\right)^n \exp\left(z n - \frac{a}{T^*}n^2\right),\qquad j\geqslant0,
	\end{equation}
	where the multiplier $1/n!$ is absent compared to~\eqref{def:K0} and~\eqref{def:Kj}.
	
	\begin{table}[h]
		\centering
		\caption{Critical point coordinates of the double-occupancy cell model for the case of indistinguishable particles. At the tricritical point, $a_{\textrm{tc}} = \frac{2}{3} \ln 2$. For any value $a \leqslant a_{\textrm{tc}}$, there is a single critical point; for any value $a > a_{\textrm{tc}}$, there are two critical points.}
		\begin{tabular}{c|c|c|c|c|c}
			\toprule
			$a=\Delta/J$ & $T^*_c$ & $\rho^{*(1)}_c $ & $P^{*(1)}_c$ & $\rho^{*(2)}_c$ & $P^{*(2)}_c$\\
			\midrule
			5.0 & $0.25 = \frac{1}{4}$ & $0.5 = \frac{1}{2}$ & $0.0482868 \approx \frac{1}{4}\left(\ln 2 - \frac{1}{2}\right)$ & $1.5 = \frac{3}{2}$ & 9.04829 \\
			2.0 & 0.250000 & 0.500000 & 0.0482868 & 1.50000 & 3.04829 \\
			1.5 & 0.250006 & 0.500018 & 0.0482895 & 1.49998 & 2.04830 \\
			1.2 & 0.250068 & 0.500204 & 0.0483169 & 1.49980 & 1.44842 \\
			1.0 & 0.250340 & 0.501020 & 0.0484374 & 1.49898 & 1.04895 \\
			0.9 & 0.250766 & 0.502302 & 0.0486268 & 1.49770 & 0.849771 \\
			0.8 & 0.251749 & 0.505276 & 0.0490662 & 1.49472 & 0.651669 \\
			0.7 & 0.254116 & 0.512503 & 0.0501341 & 1.48750 & 0.456193 \\
			0.6 & 0.260379  & 0.532171 & 0.0530373 & 1.46783 & 0.267878 \\
			0.5 & 0.283117 & 0.611865 & 0.0645817 & 1.38814 & 0.106814 \\
			$\frac{2}{3}\ln 2 \approx 0.462098$ & $\frac{1}{3} \approx 0.333333$ & 1.00000 & 0.0972532 & 1.00000 & 0.0972532\\
			0.45 & 0.378647 & 1.0 & 0.130182 & - & - \\
			0.4 & 0.452377 & 1.0 & 0.172407 & - & - \\
			0.3 & 0.532359 & 1.0 & 0.204622 & - & - \\
			0.2 & 0.587241 & 1.0 & 0.219645 & - & - \\
			0.1 & 0.630545 & 1.0 & 0.227849 & - & - \\
			0.0 & $\frac{2}{3} \approx 0.666667$  & 1.0 & $0.232408 \approx \frac{2}{3} \ln 3 - \frac{1}{2}$  & - & - \\
			\bottomrule
		\end{tabular}
		\label{tab:critical-points-ind}
	\end{table}
	
	\begin{table}[h]
		\centering
		\caption{Triple point coordinates of the double-occupancy cell model for the case of indistinguishable particles. A triple point exists in the system in the interval $a_{\textrm{tc}} \leqslant \Delta/J < 0.5$, where $a_{\textrm{tc}} = \frac{2}{3} \ln 2$.}
		\begin{tabular}{c|c|c|c|c|c}
			\toprule
			$a=\Delta/J$ & $T^*_{\textrm{tr}}$ & $P^*_{\textrm{tr}}$ & $\rho^{*(I)}_{\textrm{tr}}$ & $\rho^{*(II)}_{\textrm{tr}}$ & $\rho^{*(III)}_{\textrm{tr}}$ \\
			\midrule
			0.500 & $\approx 0.0$ & $\approx 0.0$ & $\approx 0.0$ & 1.0 & $\approx 2.0$ \\
			0.495 & 0.151204 & $6.03783 \times 10^{-3}$ & 0.0457859 & 1.0 & 1.95421 \\
			0.490 & 0.182706 & 0.0134364 & 0.0924053 & 1.0 & 1.90759 \\
			0.485 & 0.208422 & 0.0221600 & 0.145810 & 1.0 & 1.85419 \\
			0.480 & 0.232284 & 0.0324399 & 0.209923 & 1.0 & 1.79008 \\
			0.475 & 0.256113 & 0.0447467 & 0.291335  & 1.0 & 1.70867 \\
			0.470 & 0.281540 & 0.0600103 & 0.404731 & 1.0 & 1.59527 \\
			0.465 & 0.311272 & 0.0804435 & 0.600782 & 1.0 & 1.39922 \\
			$\frac{2}{3}\ln 2 \approx 0.462098$ & $0.333333 = \frac{1}{3}$ & 0.0972532 & 1.0 & 1.0 & 1.0 \\
			\bottomrule
		\end{tabular}
		\label{tab:triple-line-ind}
	\end{table}
	
	Qualitatively, the phase behavior of the model described in~sections~\ref{sec:cp}--\ref{sec:coex} remains unchanged. However, the coordinates of critical and triple points become different. These new values are given in tables~\ref{tab:critical-points-ind} and~\ref{tab:triple-line-ind}.
	Notably, at $\Delta = 0$, the critical temperature $T^*_c = \frac{2}{3}$, matching the Blume-Capel model's value from~\cite{HKBHK25}. The tricritical point is now located at $(T^*_{\textrm{tc}} = \frac{1}{3}; a_{\textrm{tc}} = \frac{2}{3} \ln 2)$, in full agreement with the Blume-Capel model result~\cite{HKBHK25}. We conclude that the proper isomorphism between the cell model (lattice gas) and the Blume-Capel model requires \emph{indistinguishable} particles to be considered.
	
	Finally, we point out that the interval of $a$, where triple points exist, narrows from $\frac{1}{2} \ln 2 < a < 0.5$ to $\frac{2}{3}\ln 2 < a < 0.5$ when changing from distinguishable to indistinguishable particles.
	
	\section{Different phases and their interpretation}
	At sufficiently large values of $a$ and at low temperatures, the double-occupancy cell model exhibits three distinct equilibrium phases labelled by I, II, and III in the order of increasing densities. At very low temperatures, these densities are $\rho^{*}_{\textrm{I}} \approx 0.0$, $\rho^*_{\textrm{II}} \approx 1.0$, and $\rho^*_{\textrm{III}} \approx 2.0$. Phase transitions between them are first-order, and the lines of these transitions terminate at critical points giving rise to phase diagrams described in section~\ref{sec:distinct}. Interestingly, the corresponding phase diagrams of the unrestricted-occupancy cell model~\cite{KD22} can contain an infinite sequence of critical points. In the present double-occupancy model, by contrast, at most two critical points occur. This suggests that restricting the maximum cell occupancy also restricts the possible number of critical points. A systematic study of this relation for general finite values of $n_{\rm max}$ is beyond the scope of the present work and is left for future investigation.
	
	Since the phases essentially differ by density, an alternative terminology can also be used. In the studies of fluids exhibiting two critical points, it is common to refer to relevant phases as gas, low-density liquid (LDL), and high-density liquid (HDL)~\cite{Franzese07}.
	We consider this terminology only as a density-based classification without implying any structural distinction between phases: our mean-field-type model cannot account for any local ordering.
	Within this interpretation, phase I corresponds to a gas, while phases II and III are LDL and HDL, respectively.
	The coexistence line between phases II and III is thus a liquid-liquid phase transition line terminating at a liquid-liquid critical point.
	
	A different terminology was used in~\cite{LYZ21}, where analogous phases were called gas, liquid, and crystal states. We avoid the term ``crystal'', since the Curie-Weiss cell model does not describe structural ordering. The high-density phase is naturally interpreted as a dense fluid phase. This point of view is supported by the behavior of the pair distribution function discussed in Appendix~\ref{sec:pair}: it resembles that of a high-density mean-field fluid~\cite{LBH00} rather than a crystalline solid.
	
	The existence of three fluid phases separated by first-order transition lines is a known feature of systems with soft-core interactions. In particular, models with attractive soft-core potentials exhibit gas, low-density liquid, and high-density liquid phases, and two critical points~\cite{Franzese07}. Similar phenomenology has been reported experimentally for several materials, including liquid phosphorus~\cite{KMUetal00,MFCM03,YKP21} and triphenyl phosphite~\cite{KT04}. Additional examples can be found in review sections of~\cite{dONCB06,Franzese07}. Despite its simplicity, the double-occupancy cell model reproduces the essential thermodynamic features associated with fluid-fluid and liquid-liquid phase transitions in systems possessing multiple density states.
	
	Although in the present work we restrict the analysis to $a \geqslant 0$, the model is well defined also for $a < 0$. In this case, the in-cell interaction becomes attractive, and double occupancy is energetically favored relative to a single occupancy. In the limit $a \to -\infty$, the singly occupied state $n_l = 1$ becomes completely suppressed, while the states $n_l = 0$ and $n_l = 2$ remain accessible. Introducing an effective occupation variable $m_l = n_l/2$, with $m_l \in \{0,1\}$, the limiting system can therefore be viewed as a single-occupancy model of particle pairs. A detailed phase behavior in the region $a < 0$ is beyond the scope of the present work and remains an interesting subject for future study.
	
	\section{Conclusions}
	
	We have studied a double-occupancy cell model of a fluid with Curie-Weiss interaction, where each cell can contain at most two particles. The interaction consists of a global attraction between all particle pairs and a local repulsion of two particles occupying the same cell. We have shown that this model is isomorphic to the Blume-Capel model on a complete graph by mapping the spin variables $S_l \in \{-1, 0, +1\}$ to occupancy numbers $n_l \in \{0, 1, 2\}$.
	We have described the thermodynamic properties and phase behavior of the model using the grand-canonical formalism previously developed for a multiple-occupancy cell model.
	
	Depending on the ratio between the repulsive and attractive couplings, the model exhibits several qualitatively distinct thermodynamic regimes. In particular, we identified regions of the phase diagram containing a single critical point, two critical points, a tricritical point, and a triple point. For sufficiently strong repulsion, three fluid phases of different densities coexist in the system, giving rise to two first-order phase-transition lines, each terminating at its own critical point. The coexistence between the intermediate-density and high-density phases may be interpreted as a liquid-liquid phase transition terminating at a liquid-liquid critical point. These results demonstrate that a remarkably rich phase behavior can emerge already in a model with the simple occupancy restriction $n_l \leqslant 2$.
	
	We considered both distinguishable and indistinguishable particles. While the phase behavior is qualitatively similar in the two cases, the formulation for indistinguishable particles reproduces the known limiting results for the Blume-Capel model on a complete graph, including the location of the tricritical point. This provides an additional confirmation of the isomorphism between the two models and indicates that the indistinguishable-particle formulation is the natural lattice-gas counterpart of the Blume-Capel model.
	
	The present model is of interest from several perspectives. As a lattice-gas-like model, it provides a simple analytically tractable framework for studying phase transitions in systems with multiple cell occupancy. From the viewpoint of soft-matter physics, it represents a minimal mean-field description of particles interacting through a competition between bounded local repulsion and long-range attraction. More broadly, the model demonstrates that the interplay between these two ingredients alone is sufficient to generate gas-liquid and liquid-liquid coexistence, two critical points, and triple points. It therefore provides a mean-field realization of a fluid exhibiting both gas-liquid and liquid-liquid phase transitions while remaining directly connected to the well-known Blume-Capel model.
	
	\section*{Acknowledgements}
	This work was supported by the National Research Foundation of Ukraine under the project No. 2023.03/0201. 
	The authors are
	deeply grateful to all warriors of the Ukrainian Armed Forces, living and fallen, for making this research possible.
	
	\appendix
	
	\section{Pair distribution function}\label{sec:pair}
	In this Appendix, we present the results for the pair distribution function $g^{(2)}$ of the double-occupancy cell model. The details of calculations can be found in~\cite{RDKPS26c}, where this quantity was obtained for the multiple-occupancy cell model. The final expression for $g^{(2)}$ as an explicit function of $T^*$ and an implicit function of $\mu^*$ is
	\begin{equation}
		\label{eq:g2_in_z}
		g^{(2)} = \frac{\left[{K}_2(T^*; \bar{z}_{\textrm{max}}) - {K}_1(T^*; \bar{z}_{\textrm{max}})\right] {K}_0(T^*; \bar{z}_{\textrm{max}})}{{K}_1(T^*; \bar{z}_{\textrm{max}})^2}.
	\end{equation}
	
	The spatial dependency of the pair distribution function is expressed as
	\begin{equation}
		\label{g2_r}
		g^{(2)}({\vb r}_1, {\vb r}_2) =  \left\{
		\begin{array}{ll}
			1, & \text{if } \nexists \Delta_l ({\vb r}_1, {\vb r}_2 \in \Delta_l),
			\\
			\frac{\langle n(n-1) \rangle_{Q}}{\langle n \rangle_{Q}^2}, & \text{if } \exists \Delta_l ({\vb r}_1, {\vb r}_2 \in \Delta_l),
		\end{array}
		\right.
	\end{equation}
	where
	\begin{equation}
		\langle n \rangle_{Q} = \rho^*,
	\end{equation}
	\begin{equation}
		\langle n(n-1) \rangle_{Q} = \sum_{n=0}^{2} n(n-1) Q_{T^*, \mu^*}(n) = 2 Q_{T^*, \mu^*}(2),
	\end{equation}
	and the probability measure $Q_{T^*,\mu^*}(n)$~\cite{KKD20} is given by:	
	\begin{equation*}
		Q_{T^*,\mu^*}(n) = \frac{\left(T^{*3/2}\right)^n}{K_0(T^*; \bar{z}_{\textrm{max}}) n!} \exp\left(\bar{z}_{\textrm{max}}n - \frac{a}{T^*}n^2 \right),
		\quad n \in \{0,1,2\},
	\end{equation*}
	in the case of distinguishable particles, and by
	\begin{equation*}
		Q_{T^*,\mu^*}(n) = \frac{\left(T^{*3/2}\right)^n}{K_0(T^*; \bar{z}_{\textrm{max}})} \exp\left(\bar{z}_{\textrm{max}}n - \frac{a}{T^*}n^2 \right),
		\quad n \in \{0,1,2\},
	\end{equation*}
	in the case of indistinguishable particles.
	
	\begin{figure}[htbp]
		\centering
		\includegraphics[width=0.5\textwidth]{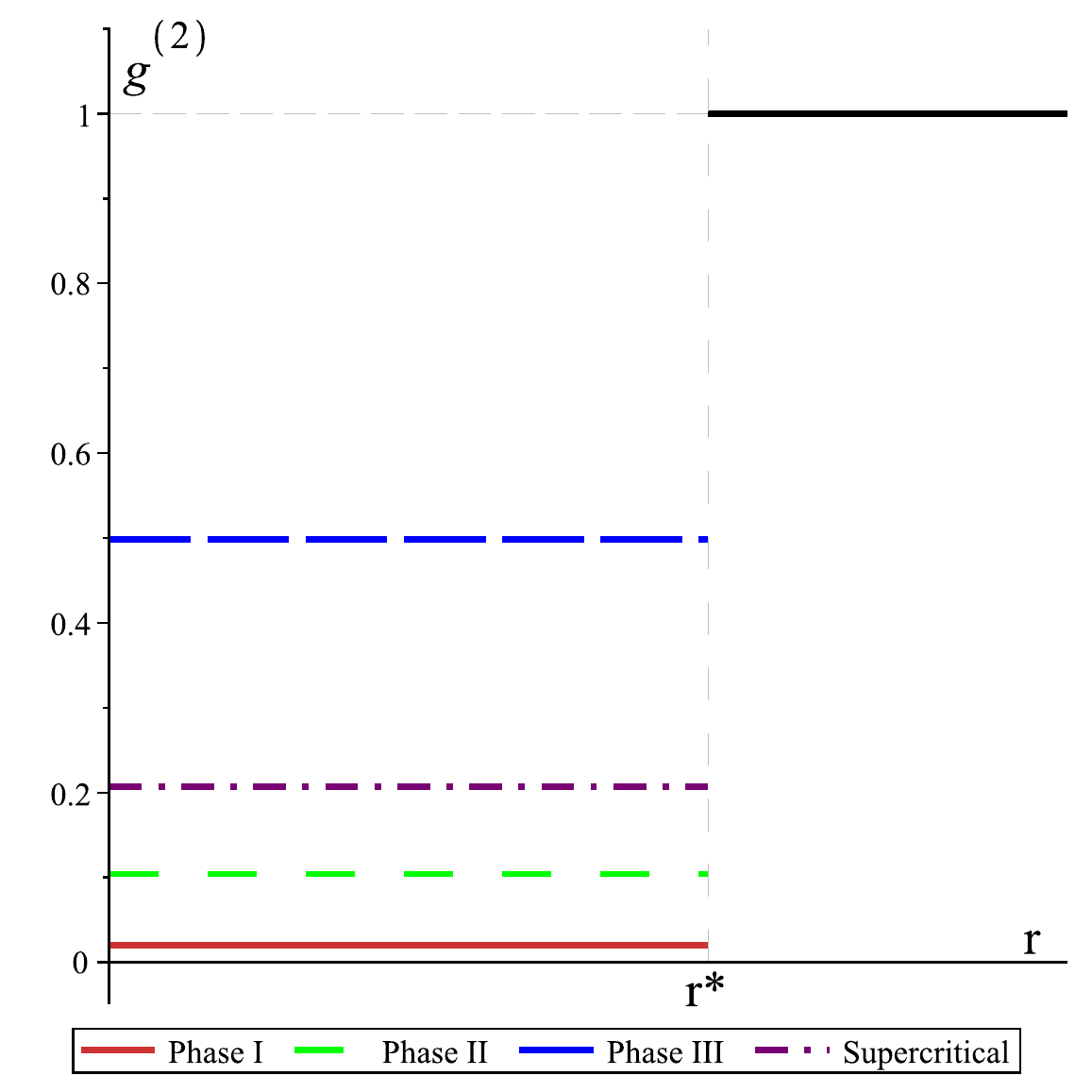}
		\caption{(Colour online) Pair distribution function $g^{(2)}({\vb r}_1, {\vb r}_2)$ as a function of separation $r = \abs{{\vb r}_1 - {\vb r}_2}$. The quantity $r^*$ denotes the distance from the point with coordinate ${\vb r}_1$ to the boundary of the cubic cell containing ${\vb r}_1$, in the direction of ${\vb r}_2$. The results are presented for the system of distinguishable particles, at $a=0.5$. Phase I: $T^* = 0.25$, $\rho^*=0.1$; Phase II: $T^* = 0.2$, $\rho^* = 1.0$; Phase III: $T^* = 0.25$, $\rho^* = 1.9$; Supercritical: $T^* = 0.5$, $\rho^* = 0.5$. At $r > r^*$, $g^{(2)} = 1$.}
		\label{fig:g2_r}
	\end{figure}
	
	In figure~\ref{fig:g2_r}, the spatial dependency of $g^{(2)} (\vb{r}_1, \vb{r}_2)$, equation~\eqref{g2_r}, is illustrated for the system of distinguishable particles at $a=0.5$ and the following selected states: for phase I, $T^*=0.25$ and $\rho^*=0.1$; for phase II, $T^*=0.2$ and $\rho^*=1.0$; for phase III, $T^*=0.25$ and $\rho^*=1.9$; and for supercritical state, $T^*=0.5$ and $\rho^*=0.5$. As is seen from the plot, $g^{(2)}$ at any selected $T^*$ and $\rho^*$ is presented as a step function. While $\vb{r}_1$ and $\vb{r}_2$ are located within one cell, $g^{(2)}$ takes on a value that is dependent on temperature and density. However, as soon as the coordinates of two particles belong to different cells, the pair distribution function becomes unity. At high densities, this is known as high-density mean field fluid, the behavior which is also observed in systems with soft interactions~\cite{LBH00}. Therefore, the result of this Appendix supports the interpretation of phase III as a dense fluid rather than a crystal.


\bibliographystyle{cmpj}

\newpage
\ukrainianpart

\title{Фазова діаграма коміркової моделі з дво-частинковим заповненням зі взаємодією Кюрі-Вейса}
%
%
\author{Р.~В.~Романік, О.~А.~Добуш, М.~П.~Козловський, І.~В.~Пилюк, М.~А.~Шпот}
\address{Iнститут фiзики конденсованих систем iменi I. Р. Юхновського НАН України, вул. Свєнціцького, 1, 79011 Львів, Україна}

\makeukrtitle

\begin{abstract}
\tolerance=3000%
Досліджено коміркову модель флюїду з подвійним заповненням та взаємодією Кюрі-Вейса. Спочатку показано, що ця модель є ізоморфною моделі Блюма-Капеля на повному графі завдяки простому перетворенню від спінових змінних до змінних заповнення. Далі її фазову поведінку досліджено в рамках великого канонічного ансамблю із застосуванням комбінації аналітичних та чисельних методів. Незважаючи на свою простоту, модель демонструє напрочуд багату термодинамічну поведінку, яка визначається співвідношенням між локальною відштовхувальною та глобальною притягальною взаємодіями. Виявлено режими, що характеризуються однією критичною точкою, двома різними критичними точками, трикритичною поведінкою та утворенням потрійної точки. За достатньо сильної відштовхувальної взаємодії система має три флюїдні фази з різними густинами, що приводить як до співіснування газової та рідинної фаз, так і до співіснування двох рідинних фаз. Визначено положення критичних, трикритичної та потрійної точок і побудовано відповідні фазові діаграми. Отримані результати демонструють, що конкуренція між відштовхуванням, пов'язаним із подвійною заселеністю комірок, та далекодійною притягальною взаємодією є достатньою для виникнення складної фазової поведінки в мінімальній ґратковій моделі газу з кратною заселеністю.
\keywords коміркова модель з подвійним заповненням, модель Блюма-Капеля, фазові переходи рідина-рідина, критична точка рідина-рідина

\end{abstract}

\end{document}